\documentclass[aps,physrev,twocolumn,superscriptaddress]{revtex4-2}

\usepackage{graphicx}
\usepackage{gensymb}
\usepackage{amsmath}
\usepackage{xcolor}

\definecolor{blue2}{RGB}{0, 0, 0}

\begin{document}

\title{Laser-generated GHz surface acoustic waves \textcolor{blue2}{with tunable amplitude} during the magnetostructural phase transition in FeRh thin films}

\author{Ia.~A.~Mogunov}
\email[]{mogunov@mail.ioffe.ru}
\affiliation{Ioffe Institute, 194021 St. Petersburg, Russia}

\author{A.~Yu.~Klokov}
\affiliation{P.N. Lebedev Physical Institute of the RAS, 119991 Moscow, Russia}

\author{N.~Yu.~Frolov}
\affiliation{P.N. Lebedev Physical Institute of the RAS, 119991 Moscow, Russia}

\author{A.~V.~Protasov}
\affiliation{Institute of Metal Physics of the Ural Branch of the RAS, 620108 Ekaterinburg, Russia}

\author{G.~E.~Zhezlyaev}
\affiliation{Institute of Metal Physics of the Ural Branch of the RAS, 620108 Ekaterinburg, Russia}

\author{D.~I.~Devyaterikov}
\affiliation{Institute of Metal Physics of the Ural Branch of the RAS, 620108 Ekaterinburg, Russia}

\author{R.~R.~Gimaev}
\affiliation{Faculty of Mechanical Engineering, University of Ljubljana, 1000 Ljubljana, Slovenia}

\author{V.~I.~Zverev}
\affiliation{Lomonosov Moscow State University, 119991 Moscow, Russia}

\author{A.~M.~Kalashnikova}
\affiliation{Ioffe Institute, 194021 St. Petersburg, Russia}

\date{\today}

\begin{abstract}

Laser-generated surface acoustic waves (SAW) facilitate efficient information processing in modern spintronics and magnonics.
The ability to tune the SAW parameters \textcolor{blue2}{such as amplitude} is crucial to achieve acoustic control over magnonic properties.
Such tunability can be achieved in phase-changing magnetic materials that accommodate both spin waves and SAWs.
A promising material is the FeRh alloy, a metallic antiferromagnet at room temperature that undergoes a phase transition to the ferromagnetic state accompanied by a crystal lattice expansion at 370~K.
This transition can also be induced by femtosecond laser pulses.
In this paper, we use the phase transition in a 60~nm Fe$_{49}$Rh$_{51}$ film to optically generate pulses of Gigahertz quasi-Rayleigh SAWs.
We detect them via the photoelastic effect and show that the lattice transformation during the phase transition is a dominant strain-generation mechanism for above-threshold excitation.
The weight of this contribution rises as the sample is heated closer to the AFM-FM transition temperature and 'switches off' when heated above it, allowing for control of the SAW amplitude.
A model based on thermodynamic parameters of Fe$_{49}$Rh$_{51}$ shows that the lattice transformation occurring within 95~ps effectively contributes to SAW generation happening on a comparable timescale, while non-equilibrium fast kinetics of the phase transition does not.

\end{abstract}

\keywords{surface acoustic waves, laser-induced phase transition, FeRh, antiferromagnet, spintronics}

\maketitle

\section{Introduction}

Spintronics and magnonics are rapidly developing as energy-efficient and versatile ways of data manipulation.
The most recent advancement constitutes antiferromagnetic spintronics \cite{AFM_spintronics_2018,AFM_spintronics_2016,Han_afm_spintronics_2023} promising unprecedentedly fast information processing at THz frequencies and information storage, insensitive to external magnetic fields.
One of the most energy efficient and versatile ways to control spins is provided by the crystal lattice and includes static deformations \cite{Bukharaev_straintronics_2018}, continuous monochromatic acoustic waves \cite{Weiler_SAW_mag_trans_2011}, standing acoustic waves \cite{Berk_mag_ac_Ni_2019}, and high-frequency strain pulses \cite{Scherbakov_1st_2010,Vlasov_review_2022,Rongione_magnetoac_NiO_2023}.
Surface acoustic waves (SAWs) have been proven to interact especially efficiently with magnons \cite{Yang_SAW_mag_review_2021,Li_SAW_mag_review_2021} allowing coherent control by establishing magnetic polarons \cite{Babu_SAW_SW_2021,Berk_mag_ac_Ni_2019} as well as other interaction regimes such as Cherenkov radiation of magnons \cite{gerevenkov2025}.
Magnetoacoustics with SAWs can provide a protected acoustic channel of communication between elements of on-chip magnonic devices \cite{Yaremkevich_guided_2021}.
Pulsed laser-generated SAWs \cite{Sugawara_ripples_SAW_2002} in tandem with laser-generated spin waves \cite{Kampen_optical_spin_waves_2002} have recently been successfully employed in magnetoacoustics due to non-contact generation, wide spectra, and high frequencies \cite{Yaremkevich_guided_2021,Babu_SAW_SW_2021}.
Magnonics and magnetoacoustics have also been proposed for unconventional computing \cite{Papp_magn_neuro_2021,Yaremkevich_reservour_2023,ac_mag_quant_2023,Whiteley_ac_mag_quant_2019}, such as neuromorphic computing \cite{Yaremkevich_reservour_2023}, which requires nonlinearities in the system \cite{neuromorph_nonlin_2019}.
One way to introduce nonlinearities into a magnetoacoustic device is to use a material with a 1$^{st}$-order phase transition \cite{Kuzum_neuro_PT_2011} between two magnetically-ordered phases.

A promising candidate for this is a FeRh alloy with 47-55~\% rhodium content, which shows a 1$^{st}$-order phase transition from antiferromagnetic (AFM) to ferromagnetic (FM) phase accompanied by the emergence of 0.9~$\mu_B$ magnetization on the Rh atom and the 1~\% isotropic volume expansion of its cubic lattice \cite{FeRh_transition_1964,Mag_atoms_1963,FeRh_content_diagram,ZVEREV_review}.
For Fe$_{49}$Rh$_{51}$ thin films, the thermally induced AFM-FM transition occurs just above room temperature in the range $T_{PT}=$360-390~K \cite{Cao_2008,Maat_2005,KOMLEV_2021,Ippei_2009}, which has led to several applications of FeRh in AFM spintronics \cite{Feng_FeRh_review,Fina_FeRh_review} and as an AFM memory cell \cite{Marti_memory_2014,Moriyama_memory_2015,Thiele_HAMR_2003}.
FeRh demonstrates a strong magnetoelastic coupling \cite{Wolloch_phonons,Lewis_2016}, allowing a magnetic-field induced AFM-FM phase transition \cite{Maat_2005} resulting in a huge volume magnetostriction of $8.2\times10^{-3}$  \cite{Ibarra_thermal_expansion,Ricodeau_magnetostr_1974}.
FeRh alloys are also well-known for their giant magnetocaloric effect \cite{ZVEREV_review,Zverev_magcal_2016,Zarkevich_magnetocal_review} and giant magnetoresistance \cite{magnetores_1995}.
The ability to trigger the AFM-FM transition in FeRh films by femtosecond laser pulses with fluence above the threshold was demonstrated in Refs.~\cite{Ju_1st_las_ind_PT_2004,Thiele_dyn_2004}.
Since then a great effort has been put into study the kinetics of magnetization \cite{Bergman_2006,Dolgikh_high_H_2025,Li_latency_2022,Mattern_8ps,Mattern_8_and_50ps,Radu_Xray_dichroism}, the crystal lattice \cite{Quirin_xray_2012,Mariager_xray_2012,Mattern_8ps,Mattern_8_and_50ps,Frazer_2021}, the spin current \cite{Kang_spin_current}, and electronic distribution \cite{Pressacco_2021} during the photoinduced phase transition (PIPT).
The main stages of the PIPT were uncovered and include sub-picosecond charge redistribution between Fe and Rh ions \cite{Pressacco_2021} followed by an 8~ps nucleation of the FM phase~\cite{Mattern_8ps,Li_latency_2022} leading to the growth and alignment of FM domains within 50-80~ps depending on experimental parameters \cite{Bergman_2006,Mariager_xray_2012,Radu_Xray_dichroism}.
The lattice expansion during PIPT follows the evolution of the volume fraction of the FM phase \cite{Mattern_8_and_50ps} with a time constant of 95~ps \cite{Quirin_xray_2012}.

The lattice change during the first-order PIPT was previously used to produce tunable high-amplitude bulk longitudinal acoustic pulses in an archetypical first-order phase transition material VO$_2$ \cite{Mogunov_gen}.
Several works investigated the generation of bulk longitudinal acoustic pulses in FeRh \cite{Quirin_xray_2012,Mattern_8ps,Kang_spin_current}, finding that they are generated and leave the excitation area on a timescale significantly shorter than that of the PIPT-related strain.
As a result, the latter provides no noticeable contribution to the amplitude of a longitudinal strain pulse \cite{Mattern_8ps,Quirin_xray_2012}.
On the other hand, \textcolor{blue2}{SAWs in FeRh films were only studied recently by Brillouin light scattering for relatively large wave vectors \cite{Ourdani_2024,ourdani_phon_mag_res_BLS} and without ultrafast PIPT excitation.
Importantly,} SAWs propagate much slower than the longitudinal acoustic waves, which poses the questions: can the lattice transformation during PIPT in FeRh be used to generate pulses of SAWs, and can their parameters be tuned by inducing the AFM-FM phase transition by external stimuli such as heating?

In this work, we generate SAW pulses directly in a FeRh film using femtosecond laser pulses and investigate their \textcolor{blue2}{amplitude} tunability by varying the external temperature and excitation fluence both below and above corresponding phase transition thresholds.
We observe SAWs in the FeRh/MgO~(001) system via the photoelastic effect as well as surface displacement and reveal a nonlinear increase of their amplitudes with increasing laser fluence when $T<T_{PT}$.
For a fixed fluence above the threshold, the generated SAW amplitude increases when the temperature approaches $T_{PT}$ from below and then drops for $T>T_{PT}$.
We devise a model attributing these findings to a pronounced contribution to SAW generation from the lattice expansion during PIPT.

\section{Experimental details}

\begin{figure}[ht]
    \includegraphics[width=1\columnwidth]{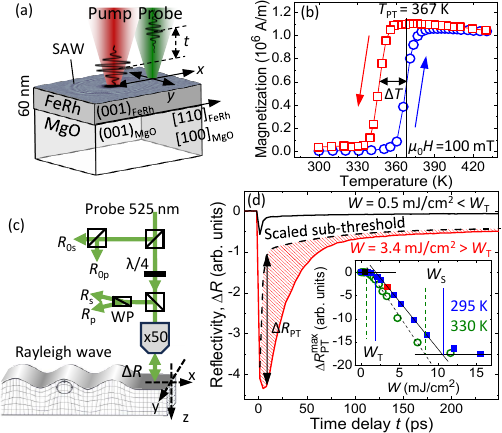}
    \caption{\label{Fig:Characterization}
    {\color{blue2}{Scheme of experiment and sample characterization.}}
    (a)~Scheme of sample and experiment;
    (b)~Thermal hysteresis of magnetization for an external in-plane field $\mu_0H$=100~mT;
    (c)~Sketch of photoelastic SAW detection.
    WP is a Wollaston prism, $\lambda$/4 is a quarter-wave plate;
    (d)~Transient reflectivity change $\Delta R(t)$ for pump fluences 0.5~mJ/cm$^2$~(black) and 3.4~mJ/cm$^2$~(red) at 295~K.
    The dotted black curve is the scaled version of the solid black one.
    The shaded area is the phase transition contribution to reflectivity $\Delta R_{PT}$. 
    Inset: fluence dependence of the maximal $\Delta R_{PT}$ value used to find the threshold and saturation fluences at 295~(blue squares) and 330~K~(green circles).
    }
\end{figure}

As a sample, we use a 60~nm epitaxial single-crystalline Fe$_{49}$Rh$_{51}$ film grown on a single-crystalline MgO~(001) substrate by molecular beam epitaxy capped with 2~nm gold layer to prevent oxidation (for details on sample growth see~\cite{KOMLEV_2021}).
The FeRh layer is grown in $(001)$ orientation confirmed by X-Ray diffraction and has a low in-plane compressive epitaxial strain $3.7\times10^{-3}$.
The surface roughness is 0.7~nm (see Supplemental Material, Sec.~I(A-B)~\cite{Suppl}).
The epitaxial relations and a schematic of the sample are shown in Fig.~\ref{Fig:Characterization}(a).
The equilibrium phase transition is characterized by the emergence of magnetization above $T_{PT}$ in an external magnetic field $\mu_0H$=100~mT along $[110]_{\text{FeRh}}$ (in-plane) exceeding the coercive field in the FM phase, $\mu_0H$=6~mT (see Supplemental Material, Sec.~I(C)~\cite{Suppl}).
We find the temperature of the AFM-FM transition to be $T_{PT}$=367~K with a width of the temperature hysteresis $\Delta T$=21~K  (Fig.~\ref{Fig:Characterization}(b)).
For FeRh films, the thermal hysteresis of magnetization matches the thermal hysteresis of its lattice parameters \cite{Arregi_growth_substrates}.

In experiments we excite and detect SAWs optically using a femtosecond pump-probe scanning microscopy where the probe pulse reflectivity change is used to detect SAWs via the photoelastic effect sensitive to various strain tensor elements including in-plane strain~\cite{Saito_method}.
The scheme of experiment is shown in Figs~\ref{Fig:Characterization}(a),(c) (see also Supplemental Material, Sec.~II~\cite{Suppl}).
Both pump and probe pulses have a duration of 150~fs and are focused by a $\times$50 micro-objective on the sample surface with normal incidence.
Pump pulses have wavelength 680~nm and are linearly polarized, probe pulses have wavelength 525~nm and are circularly polarized.
The pump and probe spots on the sample surface are circular with \textcolor{blue2}{full width at half maximum (FWHM) of} 0.9~$\mu$m and 0.7~$\mu$m, respectively.
The measured laser spots are larger than typical \textcolor{blue2}{AFM-FM phase} domains \cite{Temple_domains_2018} and grains \cite{Arregi_growth_substrates} for thin FeRh films with no external magnetic field and an order of magnitude larger than the typical size of FM phase nucleation sites \cite{Temple_domains_2018}.
The SAWs are generated in the FeRh film since the MgO substrate is optically transparent, the pump penetration depth is 20~nm \cite{Saidl_opt_params_2016}, so the pump is fully absorbed in the 60~nm-thick film.
In experiments we detect the intensity of probe pulses reflected from the sample, monitoring separately their $p$ and $s$ polarization components.
We use balanced detection with a reference beam to enhance the signal-to-noise ratio.
The time resolution in the experiments is $<$100~fs and the spatial resolution is 0.3~$\mu$m.
The sample is mounted on a heater, allowing temperature control with a 1~K precision for up to 500~K covering the range below and above $T_{PT}$=367~K (Fig.~\ref{Fig:Characterization}(b)).

\section{Results}

First, we characterize PIPT in the FeRh film in the AFM state at room temperature \textcolor{blue2}{$T_0$=}295~K and at a temperature close to the phase transition, \textcolor{blue2}{$T_0$=}330~K (Fig.~\ref{Fig:Characterization}(b)).
To do so, we measure the reflectivity change versus time delay, $\Delta R/R(t)$, for a set of pump fluences $W$.
\textcolor{blue2}{The remanent laser-induced heating does not exceed 5~K (see Supplemental Material, sec.~III \cite{Suppl}) so we consider $T_0$ to be the sample temperature.
At fluences below the threshold $W_T$, the transient reflectivity shows a fast drop followed by a slower relaxation, and $\Delta R/R$ scales linearly with fluence, which is a typical behavior for metals.
As the fluence exceeds $W_T$, the temporal evolution of the reflectivity changes due to the contribution from PIPT.}
To extract the PIPT-related reflectivity change, the scaled signal for a sub-threshold fluence is subtracted from each measured curve, an established procedure for FeRh films \cite{Mattern_8_and_50ps,Bergman_2006} \textcolor{blue2}{(see Supplemental Material sec.~III \cite{Suppl} for more detail)}.
The result is presented in Fig.~\ref{Fig:Characterization}(d) for 295~K, \textcolor{blue2}{where the shaded area highlights the contribution from PIPT, $\Delta R_{PT}$}.
We note that the reflectivity for the probe wavelength 525~nm is reduced upon PIPT, which is in accordance with the spectral changes for a temperature-driven AFM-FM transition \cite{Bennett_opt_spectra_2019,Saidl_opt_params_2016}.
The \textcolor{blue2}{maximum} PIPT-related reflectivity change, \textcolor{blue2}{$\Delta R_{PT}^{max}$}, versus pump fluence (inset in Fig.~\ref{Fig:Characterization}(d)) is clearly non-linear for both initial sample temperatures, \textcolor{blue2}{showing three distinct linear regions with boundaries at the threshold $W_T$ and saturation $W_S$ fluences.
At $W<W_T$, $\Delta R_{PT}^{max}$=0.}
At $W_T$, the PIPT starts to take place at some nucleation sites \textcolor{blue2}{within the pumped and probed volume of the FeRh film}, while above $W_S$, \textcolor{blue2}{this entire volume} undergoes the transition.
We evaluate $W_T$=2~mJ/cm$^2$, $W_S$=10.6~mJ/cm$^2$ at $T_0$=295~K and $W_T$=1.2~mJ/cm$^2$, $W_S$=8.3~mJ/cm$^2$ at $T_0$=330~K.
The values obtained are close to those reported in the literature \cite{Mattern_8ps,Bergman_2006}.

In the main experiments, we detect laser-generated SAWs and explore the PIPT contribution to the acoustic wave generation.
To do so, we fix the time delay $t$ between pump and probe pulses and scan the position of the probe spot on the sample surface, obtaining 2D maps of \textcolor{blue2}{the s-polarized component $\Delta R_s/R$ following method described elsewhere \cite{Saito_method}}.
The delay $t$=3~ns is chosen so that the SAW pulses with characteristic velocities of several km/s have enough time to leave the area excited by the pump pulse \textcolor{blue2}{(see Supplemental Material, Sec.~IV~\cite{Suppl})}.
Therefore, SAWs are detected in parts of the sample remaining in the initial state defined by the heater temperature.
We perform measurements for three initial sample temperatures -- below, close to, and above $T_{PT}$ -- and for a set of pump fluences ranging from below $W_T$ to above $W_S$.

\begin{figure}[h]
    \includegraphics[width=\columnwidth]{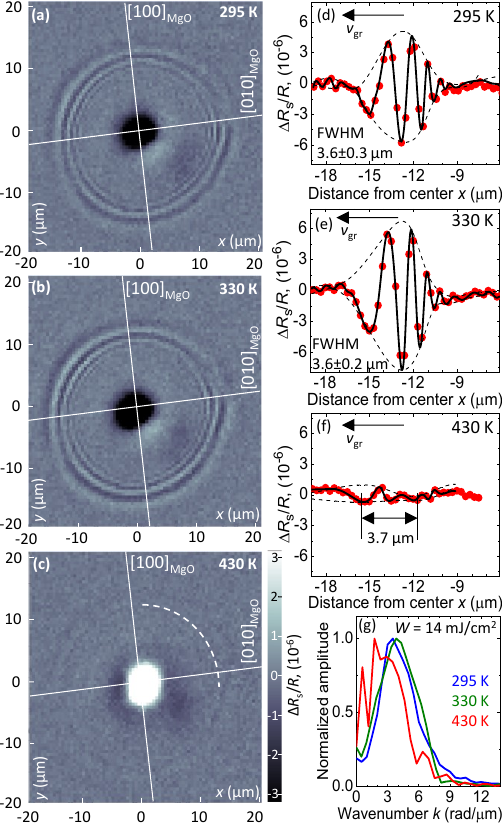}
    \caption{\label{Fig:Photoelastic_result}
    {\color{blue2}{SAW pulses in FeRh/MgO~(001) detected by the photoelastic effect below and above the phase transition temperature.}}
    (a)-(c)~2D spatial maps of $\Delta R_s/R$ for $t$=3~ns after excitation for initial sample temperatures (a)~$T_0$=295~K$<$$T_{PT}$, (b)~$T_0$=330~K$<$$T_{PT}$ and (c)~$T_0$=430~K$>$$T_{PT}$.
    The pump fluence is above saturation: $W$=10~mJ/cm$^2>W_s$.
    (d)-(f)~1D scans along the $x$ direction for the same set of temperatures as (a)-(c).
    The dots are experimental data, the solid lines are S-spline interpolations, and the dotted lines are envelopes used to determine the signal amplitudes.
    \textcolor{blue2}{Arrows indicate the SAW group velocity direction};
    (g)~Normalized FFTs of the SAW pulses from (d)-(f).
    } 
\end{figure}

Fig.~\ref{Fig:Photoelastic_result} shows exemplary 2D maps of the reflectivity change $\Delta R_s/R$ as well as their cross-sections at $y$=0 obtained at $T$=295~K, 330~K and 430~K and a pump fluence 14~mJ/cm$^2$ well above $W_S$.
In Figs.~\ref{Fig:Photoelastic_result}(a)-(c), \textcolor{blue2}{one can see a central area with decreased or increased reflectivity caused by the direct pump excitation of FeRh (the same effect for $T<T_{PT}$ can be seen in~Fig.~\ref{Fig:Characterization}(d)).
Also visible is a set of concentric, sign-alternating rings located well outside the central area (shown in more detail in Supplemental Material, Sec.~IV~\cite{Suppl})}.
The rings mark the optically detected laser-induced wave propagating away from the excitation spot with a phase velocity $v$=$\sim$5.5~km/s estimated from its shape at several time delays (see Supplemental Material, Sec.~V~\cite{Suppl}).
\textcolor{blue2}{Therefore, it outruns thermal diffusion in both FeRh phases \cite{Castellano_PRMat_2024}}.
Within experimental precision, this velocity is the same for the AFM or FM phase of FeRh.
Hence, the detected wave is not a spin wave since the dispersion of the latter changes drastically during the AFM-FM transformation \cite{CASTETS_SW_spectra_1977,Sandratskii_SW_calc}.
For a substrate loaded by a thin metallic film, one can expect three acoustic waves traveling along the surface: a quasi-Rayleigh wave, a Love wave and a surface-skimming longitudinal acoustic (SSLA) wave, with the quasi-Rayleigh wave being the slowest \cite{FARNELL197235,Sugawara_ripples_SAW_2002}.
The closest to the obtained velocity $v$ is the Rayleigh group velocity in the MgO~(001) plane, $v_R$=5.5-5.7~km/s \cite{Lee_MgO_SAW_1995}.
These two factors allow us to ascribe the observed wave to a quasi-Rayleigh SAW in the FeRh/MgO~(001) system.
The small difference between $v$ and $v_R$ is due to the mechanical loading of the MgO substrate by the FeRh film, which is a well-known feature of SAWs in a 'substrate-film' system \cite{FARNELL197235}, \textcolor{blue2}{the Rayleigh SAW velocity in FeRh is approximately 2230~m/s in the AFM phase and 2570~m/s in the FM phase, as calculated from elastic parameters \cite{Cooke_thermodyn_2012} using an approximation from Ref.~\cite{Viktorov_book_SAW}.}
Only a weak signature of the SSLA wave packet is present in the data, and no Love wave packet is observed.
Therefore, we limit further discussion to the quasi-Rayleigh wave and refer to it as 'SAW'.

Figures~\ref{Fig:Photoelastic_result}(d)-(f) show 1D scans along the $x$ direction, revealing the detailed wave packets' shape.
The solid lines show the interpolation of the data points and the dotted lines show the envelope of the interpolated data.
The FWHM of the envelope at $t$=3~ns is 3.6~$\mu$m and is the same for 295 and 330~K.
It is not possible to accurately determine the FWHM for $T_0$=430~K, but an estimation based on the signal shape yields a similar value of 3.7~$\mu$m.
A clear \textcolor{blue2}{change of frequency upon propagation (chirp)} is seen within the wave packet, caused by the dispersion of the acoustic phonons constituting the SAW pulse, as is commonly observed for laser-generated SAWs~\cite{Sugawara_ripples_SAW_2002,Kolomenskii_nonlin_SAW_2003,Klokov_diamond_2021,Tachizaki_Sagnac_2006}.
Figure~\ref{Fig:Photoelastic_result}(g) shows a fast Fourier transform (FFT) of the pulses from panels (d)-(f), normalized for clarity.
The detected SAWs possess relatively large $k$-vectors, reaching 7.5~rad/$\mu$m$^{-1}$.
The dominant $k$-vector shifts slightly to lower values for $T_0$=430~K for all the used laser fluences, although this shift cannot be confidently evaluated due to the low signal amplitude at 430~K.
The corresponding SAW wavelengths exceed 0.8~$\mu$m and are considerably larger than the FeRh film thickness of 60~nm.
Therefore, most of the acoustic energy travels in the substrate and is described by the parameters of the substrate~\cite{FARNELL197235}.

\begin{figure}[h!]
    \includegraphics[width=1\columnwidth]{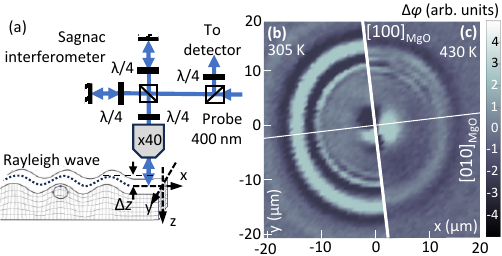}
    \caption{\label{Fig:Inteferometry_results}
    {SAW pulses in FeRh/MgO~(001) detected by interferometry below and above the AFM-FM transition temperature.}
    (a)~Sketch of the interferometer \cite{Klokov_diamond_2021} used to detect the quasi-Rayleigh SAW.
    The $\lambda$/4 components are quater-wave plates;
    (b)-(c)~2D spatial maps of $\Delta \varphi$ for $t$=3~ns after excitation with an above-threshold fluence $W$=13~mJ/cm$^2$ for initial sample temperatures (b)~$T_0$305~K$<$$T_{PT}$ and (c)~$T_0$430~K$>$$T_{PT}$.
    } 
\end{figure}

The main parameter that changes with temperature is the amplitude of the observed SAW pulses, which decreases by a factor of 8 for $T_0$=430~K.
\color{blue2}The Optical reflectivity change caused by transient strain has been studied in detail \cite{Thomsen,MATSUDA_review_2015,Saito_method}.
It consists of a complex photoelastic term (PE), which depends on combinations of the photoelastic and strain tensor components, and a purely imaginary term related to the displacement of the surface:

\begin{equation} \label{eq:refl}
    \frac{\Delta r}{r_0}=\left(\frac{\Delta r}{r_0}\right)_{\text{PE}}+\left(\frac{\Delta r}{r_0}\right)_{\text{surf}}=\left(\frac{\Delta r}{r_0}\right)_{\text{PE}}+2ik_{\text{probe}}u_0,
\end{equation}

\noindent
where $\Delta r$ is the strain-induced change of the reflection coefficient for the probe electric field, $r_0$ is its value in the unstrained medium, $k_{\text{probe}}$ is the probe wave vector, and $u_0$ is the atomic displacement along the normal at the surface ($z$=0).
In our main experiment, we measure the intensity of the reflected probe beam $\Delta R/R\approx2\text{Re}(\Delta r/r_0)$ \cite{MATSUDA_review_2015} and only sense the real part of the photoelastic term in Eq.~\ref{eq:refl}.
Therefore, this detection method is purely photoelastic, and the observed signal drop upon changing the equilibrium state of the FeRh film in Fig.~\ref{Fig:Photoelastic_result} can be caused by either a decrease in strain amplitude or a change in the photoelastic tensor. \color{black}
To verify that it is indeed the generated strain that drops above $T_{PT}$, we perform independent scanning pump-probe measurements with a different detection technique -- laser Sagnac interferometry (Fig.~\ref{Fig:Inteferometry_results}(a)) as described in \cite{Tachizaki_Sagnac_2006,Klokov_diamond_2021}, \textcolor{blue2}{a method widely used to detect quasi-Rayleigh SAW pulses \cite{Sugawara_ripples_SAW_2002,Abbas_ASOPS_SAW,Klokov_diamond_2021,Kolomenskii_nonlin_SAW_2003,Tachizaki_Sagnac_2006}.
This technique measures the phase shift of the reflected probe beam, $\varphi=\text{Im}(\Delta r/r_0)$, and is sensitive to the imaginary part of the photoelastic term as well as the full surface displacement term (see Eq.~\ref{eq:refl}).
Therefore, laser interferometry is complementary to the photoelastic detection.}
In this experiment, the laser spots on the sample are larger, 1~$\mu$m for the 800~nm probe and 1.8~$\mu$m for the 400~nm pump.

Figs.~\ref{Fig:Inteferometry_results}(b)-(c) show 2D maps of the interferometric phase difference $\Delta\varphi$ for the AFM ($T_0$=305~K) and FM ($T_0$=430~K) initial state of the FeRh film, taken 3~ns after pump excitation with an above-threshold fluence $W$=13~mJ/cm$^2$.
\textcolor{blue2}{Comparing these maps of $\Delta\varphi$ with the maps obtained previously with photoelastic detection Figs.~\ref{Fig:Photoelastic_result}(a)-(c)), one can see a similar overall picture: a central spot related to pump-induced changes and a circular SAW-related signal around it.
The detected SAW shows less pronounced chirp, which is due to the larger pump and probe spot sizes limiting the frequency resolution.
The slight 4-fold symmetry in Figs.~\ref{Fig:Inteferometry_results}(b)-(c) is attributed to the SAW velocity anisotropy within the (001) MgO plane \cite{Lee_MgO_SAW_1995}}.
Most importantly, the interferometric signal amplitude in the FM FeRh state is $\sim$4 times lower than in the AFM FeRh state, in qualitative agreement with the photoelastic results (Figs.~\ref{Fig:Photoelastic_result}(a)-(c)).
Therefore, we conclude that the laser-induced SAW amplitude decreases significantly when generated in the FM phase compared to when it is generated in the initial AFM phase.
This suggests that PIPT, which occurs only when FeRh is in its AFM state, provides a noticeable contribution to the SAW pulse generation.

\section{Discussion}\label{sec:discussion}

When considering materials with first-order PIPTs excited by femtosecond laser pulses, the strain generation process becomes non-trivial compared to conventional optoacoustic transducers \cite{Thomsen,RUELLO201521,MATSUDA_review_2015}.
This includes the material parameters change taking place during the onset of picosecond strain, as well as a direct contribution from the lattice transformation during the phase transition which can differ from the eventual structural change under quasi-equilibrium conditions~\cite{Mogunov_gen}.
The central frequency of the phonons in a SAW pulse, $f$, is a measure of the timescale of the generation process~\cite{MATSUDA_review_2015}.
The value of $f$ can be estimated from the wave vector $k$, corresponding to the maximal spectral amplitude (Fig.~\ref{Fig:Photoelastic_result}(g)), and the SAW sound velocity: $f=kv/2\pi=3.1$~GHz.
The corresponding time period $1/f$=320~ps is much longer than the duration of the main stages of the PIPT such as the 8~ps FM phase emergence~\cite{Mattern_8ps,Li_latency_2022} and the 50~ps FM phase coalescence~\cite{Bergman_2006,Mariager_xray_2012}.
Therefore, the generation of strain responsible for the SAW is regulated by material constants in the photoinduced state of FeRh, which is a mix of the initial AFM and laser-transformed FM phases, and is governed by heat diffusion \cite{Mattern_8_and_50ps} as is typically considered for laser-generated SAWs \cite{Xu_las_SAW_model,Shugaev_thermoel_SAW_gen_model}.
\textcolor{blue2}{The timescale of the lattice transformation during PIPT, 95~ps \cite{Quirin_xray_2012,Mariager_xray_2012}, while being too large to effectively contribute to coherent longitudinal phonon generation with a timescale $\sim$10~ps \cite{Mattern_8ps,Quirin_xray_2012}, it is however comparable to the 320~ps SAW period.
Therefore, we can expect a contribution from the lattice expansion to the SAW generation.}

In our experiments, we detect the SAW pulse that has already left the generation region, contrary to X-ray diffraction experiments performed within the excited area \cite{Quirin_xray_2012,Mariager_xray_2012,Mattern_8ps}.
\textcolor{blue2}{The quasi-Rayleigh SAW pulse propagating along the high-symmetry $x$ direction is elliptically polarized in the $x-z$ plane and has in-plane $\varepsilon_{xx}^{SAW}$, out-of-plane $\varepsilon_{zz}^{SAW}$, and tangent $\varepsilon_{xz}^{SAW}$ strain components.
We detect the SAW pulse within a probe penetration depth equal to 24~(29)~nm in the AFM~(FM) phase~\cite{Saidl_opt_params_2016}, which is less than the FeRh film thickness of 60~nm.
These values are much smaller than the SAW wavelength $2\pi/k\sim$1.6$\mu$m, and therefore the SAW amplitude within the probed depth can be considered constant.
This} allows us to link the strain components using boundary conditions on a free surface: \textcolor{blue2}{$\varepsilon_{xz}^{SAW}=0$,} $\varepsilon_{zz}^{SAW}/\varepsilon_{xx}^{SAW}=-\nu/(1-\nu)$ \cite{a_shameful_scitation_for_equation_for_free_boundary_strain_components} where $\nu$ is the Poisson ratio.
Using the values of $\nu$ from \cite{Cooke_thermodyn_2012}, we get $\varepsilon_{xx,AFM}^{SAW}=-2.1\varepsilon_{zz,AFM}^{SAW}$ in the AFM phase and $\varepsilon_{xx,FM}^{SAW}=-2.9\varepsilon_{zz,FM}^{SAW}$ in the FM phase.
This allows us to consider only the in-plane SAW strain component.

\begin{figure}[h]
    \includegraphics[width=\columnwidth]{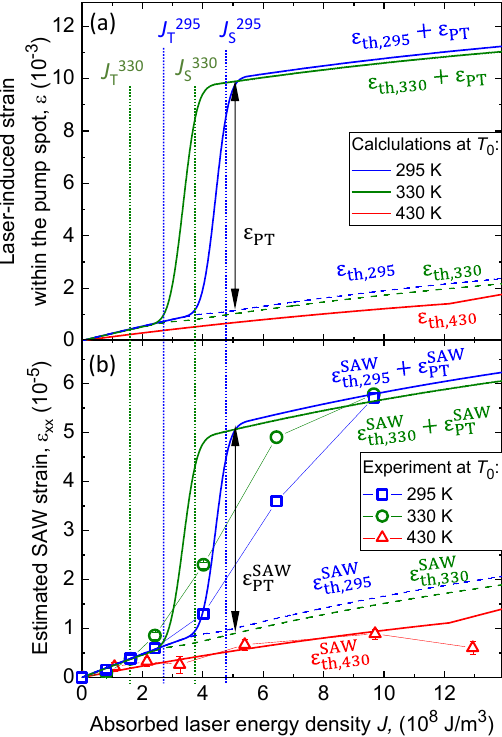}
    \caption{\label{Fig:calculation_results}
    {\color{blue2}{Laser-induced strain generation in FeRh.}}
    \textcolor{blue2}{Calculation results (lines) for initial sample temperatures $T_0$ 295~K (blue), 330~K (green) and 430~K (red).
    The dashed lines show the calculation results with $\varepsilon_{PT}=0$.
    (a)~Calculated laser-induced strain $\varepsilon$ using Eq.~\ref{eq:calc};
    (b)~Estimated in-plane SAW strain amplitudes $\varepsilon_{xx}^{SAW}$ along with the experimental data for $T_0$=295~K~(blue squares), 330~K~(green circles) and 430~K~(red triangles).}
    } 
\end{figure}

To distinguish between various contributions to the generated strain, the dependencies of \textcolor{blue2}{the maximal SAW-related reflectivity change on the} absorbed energy density $J$ are studied at three initial sample temperatures (see symbols in Fig.~\ref{Fig:calculation_results}\textcolor{blue2}{(b)}).
For our data, we calculate the values $J$ from the incident fluence $W$ taking into account the total thickness of the film $h$=60~nm rather than the pump penetration depth of 20~nm \cite{Saidl_opt_params_2016} since the SAW generation occurs slower than the heat redistribution of the laser energy deposited over the film thickness with a characteristic time of 50~ps \cite{Mattern_plasm_abs_2024,Mattern_8ps,Mattern_8_and_50ps}.
\textcolor{blue2}{The expression used is $J=(1-R)W/h$, here $R$ is the reflection coefficient in the corresponding phase of FeRh taken from Ref.~\cite{Saidl_opt_params_2016}.}
Two main conclusions can be drawn directly from the graphs in Fig.~\ref{Fig:calculation_results}\textcolor{blue2}{(b)}.
(i)~For a fixed temperature $T<T_{PT}$, the dependence of the SAW amplitude on the absorbed energy density shows a nonlinear behavior, featuring a temperature-dependent threshold at $J_T$ and saturation at $J_S$, similar to PIPT (see inset in Fig.~\ref{Fig:Characterization}(d)).
This behavior is analogous to previous reports on the longitudinal acoustic wave generated during 1st-order PIPT \cite{Mogunov_gen}.
At $T>T_{PT}$, the generated SAW exhibits a linear dependence on $J$.
(ii)~For a fixed above-threshold energy density $J>J_T$, the SAW amplitude increases with temperature until $T_{PT}$ is reached and then drops drastically.
This behavior is similar to a standalone report on SAW generation by a nanosecond pulsed laser in iron near the Curie temperature \cite{Budenkov_iron_las_SAW_1983} which was explained by a strong anomaly of the thermal expansion coefficient due to laser-heating-induced dissociation of magnetic order, causing elastic disbalance (see also \cite{Gurevich_SAW_near_FM-PM_2021}).
A similar disbalance can be caused by the lattice rearrangement during PIPT and can contribute to strain generation only below $T_{PT}$ for $J>J_T$.

In the following, we describe the mechanisms of laser-generated strain in FeRh \textcolor{blue2}{within the pump spot, $\varepsilon$,} and how we include them in our model \textcolor{blue2}{focusing on the peak strain amplitudes}.
Typically, when laser generation of SAWs in metals is analyzed, only the thermoelastic effect is considered.
The strain $\varepsilon_{th}$ is evaluated by solving the heat diffusion and thermoelastic equations together, utilizing the linear thermal expansion coefficient $\alpha$ \cite{Xu_las_SAW_model,Shugaev_thermoel_SAW_gen_model}.
For a thin film, the strain generated through the thermoelastic effect in the $z$ direction can be calculated in a simplified form as $\varepsilon_{th}(J)=\int_{T_0}^{T(J)}\alpha(T')dT'$ \cite{Thomsen,Mogunov_gen} where $T(J)$ is the laser-induced temperature.
The value of $\alpha$ for the FeRh alloy was found to drop above $T_{PT}$ and rise back above the Curie temperature $T_c=680~K$ \cite{Ibarra_thermal_expansion}.

The second mechanism is a strain $\varepsilon_{PT}$ imposed by the lattice transformation during PIPT \cite{Mogunov_gen}.
We assume that the PIPT-associated strain does not depend on the initial sample temperature if it is lower than $T_{PT}$.
The strain evolution upon PIPT in thin FeRh films on MgO~(001) was studied extensively by X-ray diffraction~\cite{Quirin_xray_2012,Mariager_xray_2012,Mattern_8ps,Grimes2022} and the lattice change in the $z$ direction was measured to be $6\times10^{-3}$ for a relatively thick FeRh film~\cite{Mattern_8ps} such as the one used in this work.
\textcolor{blue2}{However, the PIPT contribution to the laser-generated strain can differ from that measured in diffraction experiments \cite{Mogunov_gen}, so we leave $\varepsilon_{PT}$ as a free parameter.}

Finally, magnetostriction \cite{PEZERIL2016177,Bargheer_magnetostr}, spin-stress \cite{Pudell_NTE_spins,vonReppert_strain_Dy,MATTERN2023100463} and electron gas pressure \cite{RUELLO201521} can contribute to strain generation.
As we do not apply a magnetic field, the magnetic mechanisms are included in $\alpha(T)$, while from the electron gas pressure we expect no significant contribution because in FeRh the electron-phonon interaction is strong and the electrons thermalize within several hundreds of femtoseconds~\cite{Gunter_el_ph_coupl_FeRh}.
Additionally, no piezoelectric mechanism \cite{RUELLO201521} is expected because both FeRh and MgO have centrosymmetric crystal lattices.

Due to the first-order nature of the AFM-FM transition, the two phases of FeRh can coexist.
When considering the lattice change contribution, one has to account for the distribution of local $T_{PT}^*$ values of individual nucleation sites \cite{Temple_domains_2018}.
In our calculations, we assume that each nucleation site possesses a full and sharp transition at an individual $J_{PT}^*$.
This gives meaning to the values $J_T$ and $J_S$ as the boundaries of the distribution of $J_{PT}^*$, which we consider to be Gaussian, centered at $J_m=(J_T+J_S)/2$ with a dispersion $\sigma=(J_S-J_T)/6$.
The expression used to calculate the laser-induced strain $\varepsilon$ is:

\begin{multline} \label{eq:calc}
    \varepsilon(J)=\sum_{J_T<J_{PT}^*<J_S}\Bigl[\frac{\exp\left[(J_{PT}^*-J_m)^2/2\sigma^2\right]}{\sqrt{2\pi}\sigma}\\\times\Bigl( \varepsilon_{th}(J,J_{PT}^*)+\theta(J-J_{PT}^*-J_L)\varepsilon_{PT}\Bigr)\Bigr]
\end{multline}

\noindent where $\theta$ is the Heaviside function.

To calculate the thermoelastic contribution of an individual nucleation site $\varepsilon_{th}(J,J_{PT}^*)$, we first evaluate the laser-induced temperature $T$ via the absorbed energy $J$ and specific heat $C$: $J-\theta(J-J_{PT}^*-J_{L})J_{L}=\int_{T_0}^T C(T',T_{PT}^*)dT'$, where $J_{L}$=5.32~J/m$^3$ is latent heat~\cite{RICHARDSON_specific_heat} and $T_{PT}^*$ is calculated using the expression above with $J=J_{PT}^*$.
In the range between $J_{PT}^*$ and $J_{PT}^*+J_{L}$, the temperature is constant $T=T_{PT}^*$ as the phase transition is of the 1st order.
The thermoelastic contribution $\varepsilon_{th}$ is then calculated by integrating the linear thermal expansion coefficient between the initial temperature and $T$.
Both $C(T,T_{PT}^*)$ and $\alpha(T,T_{PT}^*)$ are linear approximations of the specific heat \cite{RICHARDSON_specific_heat} and the linear thermal expansion \cite{Ibarra_thermal_expansion}, respectively, above 295~K.
The discontinuities at $T_{PT}^*$ and $T_c$ \textcolor{blue2}{are introduced in place of PT-related peaks, which we explicitly take into account as latent heat and PIPT-induced lattice change} (see Supplemental Material, Sec.~VII~\cite{Suppl}).

In the calculations, the threshold value $J_T$ is defined by the nucleation of the FM phase that occurs within the penetration depth of the laser pulse.
The saturation $J_S$ energy density is defined by the slower thermal transformation through the entire depth of the FeRh film \cite{Mattern_8_and_50ps}.
This also means that $J_S/J_T\neq W_S/W_T$ since $W_T$ and $W_S$ are extracted from the change of optical properties within the probe penetration depth and possess no information about deeper regions of the FeRh film \cite{Mattern_8ps} while $J_T$ and $J_S$ are associated with the SAW generation, \textcolor{blue2}{with a localization depth $\sim$2$\pi/k$=1.6~$\mu$m containing the whole thickness of the FeRh film.}
Such definition of $J_T$ and $J_S$ allows for better agreement between the experimental and calculation results.

\textcolor{blue2}{Finally, we need to convert the laser-induced strain $\varepsilon$ into the SAW strain $\varepsilon_{xx}^{SAW}$ which is detected experimentally.
The conversion coefficients $s$ can be different for the thermoelastic $\varepsilon_{th}$ and lattice-transformation $\varepsilon_{PT}$ mechanisms since the onset of the corresponding strain happens on different timescales, and the values of $s$ can differ for FeRh in the two initial phases.
This leads to the expressions $\varepsilon_{xx}^{SAW}=s_{AFM}\varepsilon_{th}+s_{PT}\varepsilon_{PT}$ in the AFM initial phase and $\varepsilon_{xx}^{SAW}=s_{FM}\varepsilon_{th}$ in the FM initial phase of FeRh.
The values of $s$ are found by utilizing a model of an infinite stripe applying a time-dependent pressure with a Gaussian lateral distribution on a substrate \cite{Viktorov_book_SAW,Viktorov_SAW_gen_1961} (see Supplemental Material, Sec.~VI~\cite{Suppl} for details).
The obtained values are $s_{PT}\approx0.005, s_{AFM}\approx0.009$, and $s_{FM}\approx0.008$.}

The main result is shown in Fig.~\ref{Fig:calculation_results}\textcolor{blue2}{(b)} where we plot the calculation results (lines) along with the data (dots) for three initial temperatures.
The experimental SAW amplitudes are vertically scaled so that they coincide with the calculations in the region $J<J_T$.
This yields different scaling factors \textcolor{blue2}{$p=(\Delta R_s/R)/(\varepsilon_{xx}^{SAW})$ for the different initial phases: $p_{\text{AFM}}=0.35$ and $p_{\text{FM}}=0.21$}.
The scaling factors account for the possible difference in the effective photoelastic constant, as well as the change in the complex refractive index \cite{Mogunov_photoelast} and the attenuation of the SAW amplitude upon propagation for $t$=3~ns.
Note that the relation $p_{\text{AFM}}/p_{\text{FM}}\approx1.7$ accounts for the difference in the signal amplitude drop at 430~K between photoelastic (8~times, Figs.~\ref{Fig:Photoelastic_result}(d) and (f)) and interferometric (4~times, Figs~\ref{Fig:Inteferometry_results}(b) and (c)) measurements.

The calculations are based on the thermodynamic parameters of FeRh, which means that they assume thermalization of the excited sample volume.
The calculations describe the main observed features of the $T_0$- and $J$-dependencies qualitatively and capture the general shape of the $\varepsilon(J,T_0)$ curves.
The value of $\varepsilon_{PT}$ in our calculations is \textcolor{blue2}{$8.9\times10^{-3}$}, which is close to the lattice change in the $z$ direction in the FeRh/MgO films of $6\times10^{-3}$ \cite{Mattern_8ps}.
With our calculations, we have uncovered the reason for the large SAW amplitude in the AFM phase of FeRh: the large contribution from PIPT enhances the SAW amplitude in the AFM phase by a factor of 5, which can be clearly seen in Fig.~\ref{Fig:calculation_results}(b) by comparing the solid and dashed lines.
Therefore, the laser-generated SAW amplitude in FeRh is primarily defined by the AFM-FM phase transition contribution, which makes possible the tuning of the SAW \textcolor{blue2}{amplitude} by controlling the phase of FeRh.

\section{Conclusions}

In conclusion, we experimentally studied optically generated SAWs in the FeRh/MgO system.
To date, the reports on SAWs in FeRh are scarce, with only \textcolor{blue2}{three recent works \cite{SAW_in_FeRh_FMR,SAW_FeRh_transition,FeRh_SAW_neuro_2025} in which SAWs with MHz frequencies were generated elsewhere by interdigitated transducers and injected into FeRh, and a couple of works on thermal SAWs detected by Brillouin light scattering \cite{Ourdani_2024,ourdani_phon_mag_res_BLS}.}
In the present work we generated optically quasi-Rayleigh surface acoustic wave pulses with a central frequency of 3.1~GHz during the photoinduced AFM-FM phase transition in a 60~nm thick Fe$_{49}$Rh$_{51}$ film on a MgO~(001) substrate, both below and above the AFM-FM transition temperature, and for pump laser fluences ranging from subthreshold to above-saturation level of PIPT excitation.
We detected the SAWs optically via the photoelastic effect and interferometry.
The SAW amplitude for above-threshold fluences rises with temperature below $T_{PT}$ and decreases dramatically above $T_{PT}$.
For the AFM initial phase, the SAW amplitude grows nonlinearly with increasing pump fluence.
This behavior is attributed to the large contribution of the FeRh phase transition to the SAW generation process, which is present only for the AFM initial phase and above-threshold pump fluences, and is responsible for most of the laser-generated strain above threshold.
A model based on FeRh thermodynamic parameters was able to successfully describe the main features of the observed SAW amplitude dependencies on $T_0$ and $J$.
This proves that the 95~ps lattice transformation in FeRh contributes effectively to the SAW generation, while the non-equilibrium PIPT dynamics does not.

Based on our results, one can envision an optically-activated on-chip ultrafast SAW emitter in an integrated magnetoacoustic device controlled by switching between the magnetic states, AFM or FM, of FeRh \textcolor{blue2}{and operating at room temperature}.
Since FeRh was proposed earlier as a memory cell \cite{Marti_memory_2014,Moriyama_memory_2015} that stores information in the AFM state while writing is realized in the FM state, such a SAW emitter would be automatically switched off when the data is being rewritten, \textcolor{blue2}{realizing an acoustic feedback in a device.
SAWs themselves have been shown to 'write' and 'erase' the FM state in a mixed-phase FeRh film \cite{SAW_FeRh_transition}.
The utilization of such controlled acoustical cross-talk can also be used in neuromorphic computing, which was already elaborated on in Ref.~\cite{FeRh_SAW_neuro_2025} utilizing square pulses of continuous 160~MHz SAWs to demonstrate the necessary building blocks of a FeRh-based neural network.
It is an intriguing question how the SAW pulses demonstrated in our work could control a mixed-phase FeRh, since high-frequency acoustic pulses can influence phase transitions differently from quasi-static strain \cite{Mogunov_affect}.
The concept a SAW-driven neuromorphic device \cite{FeRh_SAW_neuro_2025} can benefit from non-contact (optically) generated short high-frequency SAW pulses.}

\begin{acknowledgments}

The authors thank P.~I.~Gerevenkov and Ia.~A.~Filatov for their help with the experiments and for fruitful discussions.
This research was supported by Russian Science Foundation grant No.~24-72-00111.

\end{acknowledgments}

\bibliography{FeRh_SAW}

\end{document}